\documentclass[aps,showpacs,preprintnumbers,amsmath,amssymb,nofootinbibt,preprint]{revtex4}
\usepackage{graphicx}

\begin{document}


\title{Could pressureless dark matter have pressure?}
\author{Tiberiu Harko}
\email{harko@hkucc.hku.hk} \affiliation{Department of Physics and
Center for Theoretical and Computational Physics, The University
of Hong Kong, Pok Fu Lam Road, Hong Kong, P. R. China}

\author{Francisco S.~N.~Lobo}
\email{ flobo@cii.fc.ul.pt}
\affiliation{Centro de Astronomia e Astrof\'{\i}sica da
Universidade de Lisboa, Campo Grande, Ed. C8 1749-016 Lisboa,
Portugal}

\date{\today}

\begin{abstract}

A two-fluid dark matter model, in which dark matter is represented as a two-component fluid thermodynamic system,  without interaction between the constituent particles of different species, and with each distinct component having a different four-velocity, was recently proposed in Harko \& Lobo, [Phys. Rev. D83, 124051 (2011)]. In the present paper we further investigate the two-fluid dark matter model, by assuming that the two dark matter components are pressureless, non-comoving fluids. For this particular choice of the equations of state the dark matter distribution can be described as a single anisotropic fluid, with vanishing tangential pressure, and non-zero radial pressure. We investigate the properties of this model in the region of constant velocity galactic rotation curves, where the dynamics of the test particles is essentially determined by the dark matter only. By solving the general relativistic equations of mass continuity and hydrostatic equilibrium we obtain the geometric and physical parameters of the dark matter halos in the constant velocity region in an exact analytical form. The general, radial coordinate dependent, functional relationship between the energy density and the radial pressure is also determined, and it differs from a simple barotropic equation of state.

\end{abstract}

\maketitle

\section{Introduction}

The Concordance Cosmological Model, usually referred to as the $\Lambda $ cold dark matter
($\Lambda $CDM) model, has proved to be very successful in explaining cosmological observations across a wide rage
of length scales, from the cosmic microwave background (CMB) anisotropy to the Lyman-$\alpha $
forest \cite{PeRa03,PeRa031}. In this model, nonbaryonic collisionless cold dark matter  makes up to
23\% of the total mass content of the Universe.  In the $\Lambda $CDM model, dark matter consists of cold neutral weakly interacting massive particles, beyond those
existing in the Standard Model of Particle Physics. However, up to now no dark matter candidates have been detected in
particle accelerators or in direct and indirect searches. Many particles have been proposed as
possible candidates for dark matter, the most popular ones being  the Weakly Interacting Massive Particles (WIMP) and the axions (for a review of the
particle physics aspects of dark matter see \cite{OvWe04}). The
interaction cross section of dark matter particles with normal baryonic matter is assumed to be extremely
small. However,  it is expected to be non-zero, and therefore the direct experimental detection of dark matter particles may be possible.  Superheavy particles, with mass $\geq 10^{10}$ GeV, have also been proposed as dark matter candidates. But in this case  observational results show that these
particles must either interact weakly with normal matter, or they must have masses
above $10^{15}$ GeV \cite{AlBa03}.  Scalar field models, or other long range
coherent fields coupled to gravity have also been proposed to model
galactic dark matter \cite{scal}-\cite{scal15}. The possibility that dark matter could be described by a fluid with non-zero effective pressure was also investigated \cite{pres, pres1}. In particular, it was assumed in \cite{Sax}  that the equation of state of the dark matter halos is polytropic. The fit with a polytropic dark halo improves the velocity dispersion profiles. The possibility that the galactic dynamics of
massive test particles may be understood without the need for dark matter was explored in the context of modified theories of gravity in~\cite{dmmodgrav}-\cite{dmmodgrav7}.

On galactic scales observational data  seem to disagree with the $\Lambda $CDM model predictions. High
resolution N-body simulations have shown that the predicted number of subhalos is an order
of magnitude larger than what has been observed \cite{mis}. Another discrepancy arises when
comparing the density profiles of dark halos predicted in simulations with those derived
from observations of dwarf spheroidal (dSph) galaxies and Low Surface Brightness galaxies
(LSB's).  N-body simulations predict an universal cuspy density profile \cite{Navarro:1994hi,Nav1},
\begin{equation}
\rho_{NFW}(r)\approx \frac{\rho _{s}}{(r/r_{s})(1+r/r_{s})^{2}},
\end{equation}
where $r_{s}$ is a scale
radius and $\rho _{s}$ is a characteristic density. On the other hand
observations based on high-resolution rotation curves show, instead, that the actual
distribution of dark matter is much shallower than the above, thus
indicating that a cored halo is preferred in an important fraction of low-mass galaxies \cite{Burkert:1995yz},
\begin{equation}\label{Burk}
\rho _{B}(r)\approx \frac{\rho_{0}r_{0}^{3}}{(r+r_{0})(r^{2}+r_{0}^{2})},
\end{equation}
where $r_{0}$ is the core radius
and $\rho _{0}$ is the central density. The observational Burkert density profile, given by Eq.~(\ref{Burk}), resembles an isothermal profile in the inner regions, i.e., $r\ll r_0$, predicts a finite central density, $\rho_0$, and leads to a mass profile that diverges logarithmically for increasing $r$, which is consistent with cosmological cold dark matter predictions \cite{Navarro:1994hi,Nav1}.

These discrepancies between theoretical predictions and observations might be overcome by considering other alternative dark matter models. The possibility that dark matter is a mixture of two non-interacting perfect fluids, with different four-velocities and thermodynamic parameters, was proposed recently in \cite{HaLo11}. By introducing a rotation of the four-velocity vectors the two-fluid model can be reduced to an effective single anisotropic fluid model, with distinct radial and tangential pressures \cite{Le, Le1, He}.  By assuming a non-relativistic kinetic model for the dark matter particles, the density profile and the tangential velocity of the dark matter mixture have been obtained by numerically integrating the gravitational field equations. The cosmological implications of the model have also been briefly analyzed, and it was shown that the anisotropic two-fluid model isotropizes in the large time limit. Two fluid dust models have also been considered in a general relativistic framework in \cite{Haag1, Haag2}.

It is the purpose of the present paper to further investigate the idea proposed in \cite{HaLo11}, by considering  the specific case of two, non-interacting, {\it pressureless} dark matter fluids.  For this configuration the model reduces to a single anisotropic fluid, with vanishing tangential pressure. We investigate the properties of this model in the region of constant galactic velocity rotation curves, where the solution of the basic equations can be obtained in an exact analytical form. Thus, the radial coordinate dependence of all relevant  geometric and physical parameters of the dark matter halos is explicitly determined. In particular, we obtain the general, $r$-dependent, functional relationship between the energy density and the radial pressure of the dark matter, which differs from the simple barotropic equation of state previously considered in the physical literature \cite{pres,pres1,Sax}.

The present paper is organized as follows. The two-fluid model of the dark matter halos is briefly reviewed in Section \ref{Sect2}. The general relativistic structure equations for anisotropic fluids are written down in Section~\ref{Sect3}, and the tangential velocity of test particles in stable circular orbits is obtained as a function of the geometric metric tensor. The general solution of the gravitational field equations in the constant velocity region of the dark matter halos is obtained in Section ~\ref{Sect4}. We discuss and conclude our results in Section~\ref{Sect5}.  In the present paper we use the natural system of units with $c=G=\hbar=1$.

\section{Dark matter as a mixture of two perfect fluids}\label{Sect2}

We start our study of dark matter by assuming that it consists of a mixture
of two perfect fluids, with energy densities and pressures $\rho _{1}$, $%
p_{1}$ and $\rho _{2}$, $p_{2}$, respectively, and with four velocities $%
U^{\mu }$ and $W^{\mu }$, respectively. The fluid is described by the total
energy-momentum tensor $T^{\mu \nu }$, given by
\begin{equation}
T^{\mu \nu }=\left( \rho _{1}+p_{1}\right) U^{\mu }U^{\nu }-p_{1}g^{\mu \nu
}+\left( \rho _{2}+p_{2}\right) W^{\mu }W^{\nu }-p_{2}g^{\mu \nu }.
\label{emtensor}
\end{equation}

The four-velocities are normalized according to $U^{\mu }U_{\mu }=1$ and $%
W^{\mu }W_{\mu }=1$, respectively.  The study of the physical systems
described by an energy-momentum tensor having the form given by Eq.~(\ref
{emtensor}) can be significantly simplified if we cast it into the standard
form of perfect anisotropic fluids. This can be done by means of the
transformations $U^{\mu }\rightarrow U^{\ast \mu }$ and $W^{\mu }\rightarrow W^{\ast \mu }$, respectively, so that \cite{Le,Le1, He}
\begin{equation}\label{te1}
\left(
\begin{array}{c}
U^{\ast \mu } \\
W^{\ast \mu }
\end{array}
\right) =\left(
\begin{array}{cc}
\cos \alpha  & \sqrt{\frac{\rho _{2}+p_{2}}{\rho _{1}+p_{1}}}\sin \alpha  \\
-\sqrt{\frac{\rho _{1}+p_{1}}{\rho _{2}+p_{2}}}\sin \alpha  & \cos \alpha
\end{array}
\right) \left(
\begin{array}{c}
U^{\mu } \\
W^{\mu }
\end{array}
\right) ,
\end{equation}
representing a ``rotation'' of the velocity four-vectors in the $\left(
U^{\mu },W^{\mu }\right) $ velocity space. Notice that the transformations
given by Eqs.~(\ref{te1})  leave the quadratic form $\left(
\rho _{1}+p_{1}\right) U^{\mu }U^{\nu }+\left( \rho _{2}+p_{2}\right) W^{\mu
}W^{\nu }$ invariant. Thus we have
\begin{equation}
T^{\mu \nu }\left( U,W\right) =T^{\mu \nu
}\left( U^{\ast },W^{\ast }\right) .
\end{equation}
 As for the vectors  $U^{\ast \mu}$ and $W^{\ast \mu}$ we assume
that one is timelike, while the other is spacelike, so that
$U^{\ast \mu }W_{\mu }^{\ast }=0$.

With the use of the latter relationship and Eqs.~(\ref{te1}) we obtain the rotation angle as
\begin{equation}\label{angle}
\tan 2\alpha =2\frac{\sqrt{\left( \rho _{1}+p_{1}\right) \left( \rho
_{2}+p_{2}\right) }}{\rho _{1}+p_{1}-\left( \rho _{2}+p_{2}\right) }U^{\mu
}W_{\mu }.
\end{equation}

 Next we define the following quantities \cite{Le,Le1,He}:
\begin{equation}
V^{\mu }=\frac{U^{\ast \mu }}{\sqrt{U^{\ast \alpha }U_{\alpha }^{\ast }}}%
,\qquad \chi ^{\mu }=\frac{W^{\ast \mu }}{\sqrt{-W^{\ast \alpha }W_{\alpha }^{\ast }%
}},
\end{equation}
\begin{equation}
\varepsilon =T^{\mu \nu }V_{\mu }V_{\nu }=\left( \rho _{1}+p_{1}\right)
U^{\ast \alpha }U_{\alpha }^{\ast }-\left( p_{1}+p_{2}\right) ,
\end{equation}
\begin{equation}
\sigma =T^{\mu \nu }\chi _{\mu }\chi _{\nu }=\left( p_{1}+p_{2}\right)
-\left( \rho _{2}+p_{2}\right) W^{\ast \alpha }W_{\alpha }^{\ast },
\end{equation}
\begin{equation}
\Pi =p_{1}+p_{2},
\end{equation}
respectively. Thus, the energy-momentum tensor of the two non-interacting perfect fluids
can be written as
\begin{equation}
T^{\mu \nu }=\left( \varepsilon +\Pi \right) V^{\mu }V^{\nu }-\Pi g^{\mu \nu
}+\left( \sigma -\Pi \right) \chi ^{\mu }\chi ^{\nu },  \label{tens}
\end{equation}
where
\begin{equation}
V^{\mu }V_{\mu }=1=-\chi ^{\mu }\chi _{\mu },
\end{equation}
and
\begin{equation}
\chi ^{\mu }V_{\mu }=0.
\end{equation}\texttt{}
Note that the energy-momentum tensor given by Eq.~(\ref{tens}) is the standard form for anisotropic fluids \cite{He}.

The energy density $\varepsilon $ and the radial pressure $\sigma $ are
given by
\begin{eqnarray}
\varepsilon &=&\frac{1}{2}\left( \rho _{1}+\rho _{2}-p_{1}-p_{2}\right) +\nonumber\\
&&\frac{%
1}{2}\sqrt{\left( \rho _{1}+p_{1}+\rho _{2}+p_{2}\right) ^{2}+4\left( \rho
_{1}+p_{1}\right) \left( \rho _{2}+p_{2}\right) \left[ \left( U^{\mu }W_{\mu
}\right) ^{2}-1\right] },  \label{eps}
\end{eqnarray}
and
\begin{eqnarray}
\sigma &=&-\frac{1}{2}\left( \rho _{1}+\rho _{2}-p_{1}-p_{2}\right) +\nonumber\\
&&\frac{1}{2%
}\sqrt{\left( \rho _{1}+p_{1}-\rho _{2}-p_{2}\right) ^{2}+4\left( \rho
_{1}+p_{1}\right) \left( \rho _{2}+p_{2}\right) \left( U^{\mu }W_{\mu
}\right) ^{2}},  \label{sig}
\end{eqnarray}
respectively \cite{Le,Le1,He}.

In comoving spherical coordinates $x^{0}=t$, $x^{1}=r$, $%
x^{2}=\vartheta $, and $x^{3}=\phi $ we may choose $V^{1}=V^{2}=V^{3}=0$, $%
V^{0}V_{0}=1$, and $\chi ^{0}=\chi ^{2}=\chi ^{3}=0$, $\chi ^{1}\chi _{1}=-1$ \cite{Le,Le1,He}. Therefore the components of the energy-momentum of two non-interacting
perfect fluids take the form
\begin{equation}
T_{0}^{0}=\varepsilon , T_{1}^{1}=-\sigma, T_{2}^{2}=T_{3}^{3}=-\Pi ,
\label{eqs}
\end{equation}
where $\varepsilon $ is the total energy-density of the
mixture of fluids, $\sigma =P_{r}$ is the pressure along the radial
direction, while $\Pi =P_{\perp }$ is the tangential pressure on the $r=$
constant surface.

\section{Anisotropic fluids in spherically symmetric static spacetimes}\label{Sect3}

In the following we restrict our study of the two-component dark matter to
the static and spherically symmetric case, with the metric  represented as
\begin{equation}\label{metr1}
ds^{2}=e^{\nu (r)}dt^{2}-e^{\lambda (r)}dr^{2}-r^{2}\left( d\vartheta
^{2}+\sin ^{2}\vartheta d\phi ^{2}\right) .
\end{equation}

For the metric given by Eq.~(\ref{metr1}), the Einstein gravitational field
equations, describing the mixture of two pressureless fluids, take the form~
\begin{eqnarray}
-e^{-\lambda }\left( \frac{1}{r^{2}}-\frac{\lambda ^{\prime }}{r}\right) +%
\frac{1}{r^{2}}=8\pi \varepsilon ,  \label{f1} \\
e^{-\lambda }\left( \frac{\nu ^{\prime }}{r}+\frac{1}{r^{2}}\right) -\frac{1%
}{r^{2}}=8\pi \sigma ,  \label{f2}\\
\frac{1}{2}e^{-\lambda }\left( \nu ^{\prime \prime }+\frac{\nu ^{\prime 2}}{2%
}+\frac{\nu ^{\prime }-\lambda ^{\prime }}{r}-\frac{\nu ^{\prime }\lambda
^{\prime }}{2}\right) =8\pi \Pi ,  \label{f3}
\end{eqnarray}
and
\begin{equation}
\nu ^{\prime }=-2\frac{\sigma ^{\prime }}{\varepsilon +\sigma }+\frac{4}{r}%
\frac{\Pi -\sigma }{\varepsilon +\sigma },  \label{f4}
\end{equation}
where $^{\prime }=d/dr$. Equation (\ref{f4}) is the consequence of the
conservation of the energy-momentum tensor, $T^{\mu }{}_{\nu;\mu }=0$. Eq.~(\ref{f1}) can be easily integrated to give
\begin{equation}
e^{-\lambda }=1-\frac{2M(r)}{r},
\end{equation}
where $M(r)=4\pi \int \varepsilon r^{2}dr$. W\texttt{}ith the use of Eqs.~(\ref{f2}) and (\ref{f4}) we obtain the mass
continuity equation, and the generalized Tolman-Oppenheimer-Volkoff (TOV)
equation describing the static pressureless two-fluid dark matter
configurations,
\begin{equation}
\frac{dM}{dr}=4\pi \varepsilon r^{2},  \label{eqcont0}
\end{equation}
\begin{equation}
\frac{d\sigma }{dr}=-\frac{\left( \varepsilon +\sigma \right) \left( 4\pi
\sigma r^{3}+M\right) }{r^{2}\left( 1-2M/r\right) }+\frac{2}{r}\left( \Pi
-\sigma \right) .  \label{tov0}
\end{equation}

The galactic rotation curves provide the most direct method of analyzing the
gravitational field inside a spiral galaxy. The rotation curves are obtained
by measuring the frequency shifts $z$ of the 21-cm radiation emission from
the neutral hydrogen gas clouds. Usually astronomers report the
resulting $z$ in terms of a velocity field $v_{tg}$ \cite{dm,dm1,dm2, dm3}. The line
element given by Eq.~(\ref{metr1}) can be rewritten in terms of the spatial
components of the velocity, normalized with the speed of light, measured by
an inertial observer far from the source, as $ds^{2}=dt^{2}\left(
1-v^{2}\right) $, where
\begin{equation}
v^{2}=e^{-\nu }\left[ e^{\lambda }\left( \frac{dr}{dt}\right)
^{2}+r^{2}\left( \frac{d\Omega }{dt}\right) ^{2}\right] .
\end{equation}

For a stable circular orbit $dr/dt=0$, the tangential velocity of the
test particle can be expressed as
\begin{equation}
v_{tg}^{2}=e^{-\nu }r^{2}\left( \frac{d\Omega }{dt}\right) ^{2},
\end{equation}
where $d\Omega ^2=d\vartheta ^{2}+\sin ^{2}\vartheta d\phi ^{2}$.

In terms of the conserved angular momentum $l=r^{2}\dot{\phi}$ and energy $E=e^{\nu (r)}\dot{t}$,
the tangential velocity is given, for $\theta =\pi /2$, by
\begin{equation}
v_{tg}^{2}=\frac{e^{\nu }}{r^{2}}\frac{l^{2}}{E^{2}}.
\end{equation}

By taking into account the explicit expressions for $l$ and $E$ we obtain
for the tangential velocity of a test particle in a stable circular orbit
the expression
\begin{equation}
v_{tg}^{2}=\frac{1}{2}r\nu ^{\prime }.  \label{vtg}
\end{equation}

The tangential velocity profile can be expressed as a function of the total
density and of the mass of the dark matter from Eq.~(\ref{f2}) as
\begin{equation}
v_{tg}^{2}(r)=\frac{4\pi \sigma r^{3}+M}{r\left( 1-2M/r\right) }.
\end{equation}

The TOV Eq.~(\ref{tov0}), can be written in the following equivalent form
\begin{equation}
\frac{d\sigma }{dr}=-\frac{\left( \varepsilon +\sigma \right) }{r}%
v_{tg}^{2}(r)+\frac{2}{r}\left( \Pi -\sigma \right) .  \label{tovf}
\end{equation}

\section{Dark matter as a mixture of two pressureless fluids}\label{Sect4}

In order to describe dark matter we adopt the kinetic model considered in \cite{HaLo11}. The  energy-momentum tensor $T^{\mu }_{\nu }$ associated to the frozen distribution of dark matter is given by $T^{\mu }_{\nu }=g\int{d^3pf(p)p^{\mu }p_{\nu }/p^0}$, where $f(p)$ is the dark matter particle distribution function, $p^{\mu }$ is the four-momentum,  $\vec{p}$ is the three-momentum, with absolute value $p$,  and $g$ is the number of helicity states, respectively.
The energy density $\epsilon $ of the system is defined as
\begin{equation}
\epsilon =\frac{g}{3}\int{Ef(p)d^3p},
\end{equation}
while the pressure of a system with an isotropic distribution of momenta is given by
\begin{equation}
P=\frac{g}{3}\int{pvf(p)d^3p}=\frac{g}{3}\int{\frac{p^2}{E}f(p)d^3p},
\end{equation}
where the velocity $v$ is related to the momentum by $v=p/E$ \cite{mad, mad1}.
In the non-relativistic regime, when $E\approx m$ and $p\approx mv$,   the density $\rho $ of the dark matter is given by $\rho =mn$, where $n$ is the particle number density, while its pressure $P$ can be obtained as \cite{mad, mad1}
\begin{equation}\label{pres0}
P=\frac{g}{3}\int{\frac{p^2}{E}f(p)d^3p}\approx 4\pi \frac{g}{3}\int{\frac{p^4}{m}dp},
\end{equation}
giving
\begin{equation}\label{pres1}
P=\frac{ \langle \vec{v}^{\;2} \rangle }{3}\rho ,
\end{equation}
where $\langle \vec{v}^{\;2} \rangle$ is the average squared velocity of the particle, and $\langle \vec{v}^{\;2} \rangle /3$ is the one-dimensional velocity dispersion. In the non--relativistic approximation given by Eqs.~(\ref{pres0}) and (\ref{pres1}), the velocity dispersion  is  a constant only for the
case of the non--degenerate ideal Maxwell--Boltzmann gas whose Newtonian analogue is
the isothermal sphere. In the following we will consider only the specific case of the ideal gases in the non--relativistic regime. Since for non-relativistic particles $\langle \vec{v}^{\;2} \rangle/3 \ll 1$, and their energy density is also very small, we can consider, with a very good approximation, that each component of the dark matter is pressureless, $P\approx 0$.
Therefore we assume that dark matter
consists of the mixture of two pressureless fluids, so that $p_{1}=p_{2}\approx 0$.
The two fluids have different four-velocities, and hence we may write
\begin{equation}
U^{\mu }W_{\mu }=1+\frac{b}{2},
\end{equation}
where in the general case $b$ is an arbitrary function of the radial coordinate $r$. The functional form of $b$ can be estimated by using Eq.~(\ref{angle}). If the angle $\alpha $ between the two four-velocities  is small, we obtain
\begin{equation}
U^{\mu }W_{\mu }\approx \frac{\rho _1-\rho _2}{\sqrt{\rho _2/\rho _1}}\alpha .
\end{equation}
In the following we assume that the number density of the two components is roughly the same, $n_1=n_2=n$, but the dark matter particles have different masses $m_1\neq m_2$. Thus $\rho _2/\rho _1=m_2/m_1$, and $b$ can be estimated as
\begin{equation}
b\approx 2\left(\frac{1-m_2/m_1}{\sqrt{m_2/m_1}}\alpha -1\right).
\end{equation}

The components of the effective anisotropic
energy-momentum tensor for this mixture of fluids is given by
\begin{equation}
\varepsilon =\frac{1}{2}\left( \rho _{1}+\rho _{2}\right) +\frac{1}{2}\sqrt{%
\left( \rho _{1}+\rho _{2}\right) ^{2}+\left( b^{2}+4b\right) \rho _{1}\rho
_{2}},
\end{equation}
\begin{equation}
\sigma =P_{r}=-\frac{1}{2}\left( \rho _{1}+\rho _{2}\right) +\frac{1}{2}%
\sqrt{\left( \rho _{1}-\rho _{2}\right) ^{2}+\left( b+2\right) ^{2}\rho
_{1}\rho _{2}},
\end{equation}
\begin{equation}
\Pi =P_{\bot }=0,
\end{equation}
respectively. Therefore the mixture of the two pressureless fluids can be
modeled as an anisotropic fluid with vanishing tangential pressure. In the
following we will restrict our study to the constant tangential velocity
region, so that we can take $v_{tg}=$ constant. This assumption determines,
with the use of Eq.~(\ref{vtg}), the functional form of the metric tensor
component in the dark matter region as
\begin{equation}
e^{\nu }=\left( \frac{r}{r_{0\nu }}\right) ^{2v_{tg}^{2}},
\end{equation}
where $r_{0\nu }$ is an arbitrary integration constant. The gravitational
field equation Eq.~(\ref{f3}) can be written as
\begin{equation}
\lambda ^{\prime }=\frac{2v_{tg}^{2}}{1+v_{tg}^{2}}\frac{1}{r},
\end{equation}
which immediately gives
\begin{equation*}
e^{\lambda }=\left( \frac{r}{r_{0\lambda }}\right) ^{2v_{tg}^{2}/\left(
1+v_{tg}^{2}\right) }.
\end{equation*}
where $r_{0\lambda }$ is an arbitrary integration constant. The density
profile of the two component dark matter mixture follows from Eq.~(\ref{f1}%
), and is given by
\begin{equation}\label{prof}
8\pi \varepsilon =\frac{1}{r^{2}}\left[ 1-\frac{1-v_{tg}^{2}}{1+v_{tg}^{2}}%
\left( \frac{r}{r_{0\lambda }}\right) ^{-{2v_{tg}^{2}/\left(
1+v_{tg}^{2}\right) }}\right] ,
\end{equation}
while the effective pressure of the two component dark matter is given by
\begin{equation}
8\pi \sigma =P_{r}=\frac{1}{r^{2}}\left[ \left( 1+2v_{tg}^{2}\right) \left(
\frac{r}{r_{0\lambda }}\right) ^{-2v_{tg}^{2}/\left( 1+v_{tg}^{2}\right) }-1%
\right] .  \label{press}
\end{equation}

Equation (\ref{press}) enables us to introduce the radius $R$ of the dark matter
distribution, defined as the vacuum boundary of the system, where the radial
pressure vanishes, $\sigma \left( R\right) =0$. This condition gives
\begin{equation}
r_{0\lambda }=\left( 1+2v_{tg}^{2}\right) ^{-\left( 1+v_{tg}^{2}\right)
/2v_{tg}^{2}}R.
\end{equation}

Thus, the mass profile of the dark matter is provided by
\begin{equation}
m(r)=\frac{1}{2}\left[ r-r_{0\lambda }\left( \frac{r}{r_{0\lambda }}\right)
^{\left( 1-v_{tg}^{2}\right) /\left( 1+v_{tg}^{2}\right) }\right] .
\end{equation}
The total mass of the dark halo is given by $M=m(R)$, and, by using the
relation
\begin{equation}
\left( \frac{R}{r_{0\lambda }}\right) ^{\left( 1-v_{tg}^{2}\right) /\left(
1+v_{tg}^{2}\right) }= \frac{1}{ 1+2v_{tg}^{2}}  \left(
\frac{R}{r_{0\lambda }}\right) ,
\end{equation}
can be obtained in a simple form as
\begin{equation}
M=\frac{v_{tg}^{2}R}{(1+2v_{tg}^{2})}.
\end{equation}
The dark matter density at the boundary surface is given by
\begin{equation}
8\pi \varepsilon \left( R\right) =\frac{1}{R^{2}}\left[ 1-\frac{1-v_{tg}^{2}%
}{\left( 1+v_{tg}^{2}\right) \left( 1+2v_{tg}^{2}\right) }\right] .
\end{equation}

The equation of state of dark matter can be formally obtained from Eq. (%
\ref{tovf}). Since $\Pi =0$, Eq.~(\ref{tovf}) can be written as a first
order differential equation,
\begin{equation}
\frac{d\sigma }{dr}=-\frac{2+v_{tg}^{2}}{r}\sigma -v_{tg}^{2}\frac{%
\varepsilon }{r},
\end{equation}
which can be integrated to give
\begin{equation}
\sigma \left( r\right) =r^{-\left( 2+v_{tg}^{2}\right) }\left[ \sigma
_{0}+\int r^{1+v_{tg}^{2}}\varepsilon \left( r\right) dr\right] ,
\end{equation}
where $\sigma _0$ is an arbitrary integration constant.

However, in the present model the equation of state is not of a simple
barotropic form, and the functional relation between $\varepsilon $ and $%
\sigma $ is also $r$-dependent. Once $\varepsilon $ and $\sigma $ are
known, from Eqs.~(\ref{eps}) and (\ref{sig}) one can obtain the density
profiles of the two dark matter components, which are given by
\begin{equation}
\rho _{i}=\frac{1}{2}\left[ \epsilon -\sigma \pm \sqrt{{\epsilon }^{2}-\frac{%
2\,\left( 8+b\,\left( b+4\right) \right) \epsilon \sigma }{b\left(
b+4\right) }+{\sigma }^{2}}\right] , i=1,2.
\end{equation}

\section{Discussions and final remarks}\label{Sect5}

In the present paper we have further considered the theoretical possibility, proposed in \cite{HaLo11},  that dark matter may be modeled as a mixture of two non-interacting perfect fluids, with different four-velocities. The most interesting characteristic of  this model is that it is formally equivalent to a single anisotropic fluid. It is important to note that this equivalence is general, and it is independent of the stationarity condition of the metric assumed in the present paper. Mathematically, this result follows from the invariance of the total energy-momentum tensor of a two-fluid system with respect to the group of rotations in the velocity space. Thus, the same approach can be used for the study of time-dependent dark matter halos, like, for example, those resulting from the collision of two galaxies.   In particular, for the specific case of two non-interacting pressureless fluids, the two fluid model reduces to a single anisotropic fluid with vanishing tangential pressure.

By considering the flat galactic rotation curves region, we have completely solved the general relativistic structure equations of the pressureless two fluid dark matter halos. We have  found a general, $r$-dependent functional relationship between the energy density and the radial pressure of the dark matter halos, which differs radically from the simple barotropic equation of state previously considered in the literature. On the other hand, in was shown in \cite{Sax} that half of the 14 galaxies considered in the study are well fit by the polytropic halo model, despite its serious physical simplifications.

The necessity of considering  non-standard dark matter models with pressure is justified by our uncertainties in the knowledge about the nature of the dark matter particles, as well as by the fact that this model gives a much better description of the observational results, as compared to the pressureless case. On the other hand, the very successful standard $\Lambda $CDM cosmological model is based on the fundamental assumption of the existence of the pressureless dark matter.

The apparent contradiction between these two approaches could be solved by assuming that dark matter is a mixture of two pressureless fluids, with different four-velocities. Then dark matter can be described as a single, anisotropic fluid. Similarly to the case of the NFW profile, the anisotropic dark matter profile given by Eq.~(\ref{prof}) diverges for $r\rightarrow 0$, thus predicting a cuspy profile. However, this unphysical divergence of the density is the result of our assumption of a constant tangential velocity, which enables us to obtain a complete analytical solution of the full set of the gravitational field equations. By assuming an arbitrary velocity profile, the field equations can be solved numerically, as done in \cite{HaLo11}, and the obtained density distributions have a non-singular, core-like profile.  From an analytical point of view, one way to deal with this unphysical feature would be to perform a smooth matching between the anisotropic fluid distribution, and a small central region with a regular density profile.  This approach would lead to a better analytical description
of the dark matter halos. Another possibility would be to assume that indeed each of the two components of the dark matter fluid have some pressure \cite{HaLo11}, and consider some approximate solutions of the field equations. Some of these possibilities, as well as their impact on the dark matter distribution and properties, together with the cosmological implications, will be considered in a future study.

\section*{Acknowledgments}

We would like to thank the anonymous referee for comments and suggestions that helped us to significantly improve our manuscript. The work of TH was supported by an GRF grant of the government of the Hong Kong SAR. FSNL acknowledges financial support of the Funda\c{c}\~{a}o para a Ci\^{e}ncia e Tecnologia through the grants PTDC/FIS/102742/2008 and CERN/FP/116398/2010.

\end{document}